\documentclass[12pt]{article}
\usepackage{kotex}
\usepackage[round]{natbib}
\usepackage{graphicx}
\usepackage{amssymb}
\usepackage{pdflscape}
\usepackage{hyperref}
\usepackage{tikz}
\usepackage{amsmath}
\usepackage{enumitem}
\usepackage{array}
\usepackage{longtable}
\usepackage{booktabs}
\usepackage{afterpage}
\usepackage[margin=1in]{geometry}
\usepackage{authblk}

\usetikzlibrary{shapes.geometric, arrows}
\hypersetup{colorlinks,linkcolor=[RGB]{223 106 106},citecolor={blue},urlcolor={blue}}

\newcommand{\Nincl}{18}
\newcommand{\Nexcl}{7}
\newcommand{\Ntotal}{25}

\begin{document}

\title{Visualization of AI Systems in Virtual Reality: A Comprehensive Review}
\author[1]{Medet Inkarbekov\thanks{Contact Medet Inkarbekov. Email: medet.inkarbekov@mu.ie}}
\author[1,2]{Rosemary Monahan}
\author[1,2]{Barak A. Pearlmutter}

\affil[1]{Department of Computer Science, Maynooth University, Maynooth, Ireland}
\affil[2]{Hamilton Institute, Maynooth University, Maynooth, Ireland}

\maketitle

\begin{abstract}
  This study provides a comprehensive review of the utilization of Virtual Reality (VR) for visualizing Artificial Intelligence (AI) systems, drawing on 18 selected studies. The results illuminate a complex interplay of tools, methods, and approaches, notably the prominence of VR engines like Unreal Engine and Unity. However, despite these tools, a universal solution for effective AI visualization remains elusive, reflecting the unique strengths and limitations of each technique. We observed the application of VR for AI visualization across multiple domains, despite challenges such as high data complexity and cognitive load. Moreover, it briefly discusses the emerging ethical considerations pertaining to the broad integration of these technologies. Despite these challenges, the field shows significant potential, emphasizing the need for dedicated research efforts to unlock the full potential of these immersive technologies. This review, therefore, outlines a roadmap for future research, encouraging innovation in visualization techniques, addressing identified challenges, and considering the ethical implications of VR and AI convergence.
\end{abstract}

\section{Introduction}
\label{sec:introduction}

Artificial intelligence (AI) systems have advanced quickly in recent years, with applications ranging from robotics and autonomous vehicles to machine learning and natural language processing \citep{seshia2022}. As these systems become more complex, effective visualisation methods are becoming increasingly important \citep{chatzimparmpas2020}.  Primarily, it fosters an understanding of AI operations, demystifying the 'black box' nature of these systems into interpretable processes \citep{liang2021}. This increased transparency aids not only in understanding the system's operations but also in spotting errors or inefficiencies, thus, paving the path for necessary improvements. Visualization also plays a pivotal role in enhancing the trust and transparency of AI systems, by illuminating the decision-making pathways of these technologically advanced systems \citep{schmidt2020}. Lastly, visualization is a key player in education \citep{samek2017}, converting abstract AI concepts into tangible, clear forms. Thus, it stands as an indispensable tool in making intricate AI models more understandable, trustworthy, and user-friendly.

Virtual reality (VR), with its immersive and interactive capabilities, offers a promising platform for visualizing AI systems \citep{Korkut2023}. Users can experience and interact with AI systems in an immersive and interactive environment via VR, and visualization techniques can aid users in better grasping and analyzing the data produced by AI systems.

The immersive experience of VR can offer new possibilities to make Convolutional Neural Networks (CNNs) more human-understandable by depicting them in 3D environments \citep{Schreiber2019}. However, according to \citet{Linse2022}, this field is still relatively unexplored and has two main challenges. First, rendering large, popular architectures like ResNet50 is currently not feasible with existing tools, which often restrict CNNs to linear structures without splits or joints. Additionally, the number of visible layers may be limited due to computational limitations or constraints with the interaction design. Second, current tools do not offer a flexible and convenient interface for developers and researchers to visualize custom architectures. To fully realize the potential of VR for understanding and interacting with CNNs, more research and development are needed in this field. Despite the potential benefits of visualizing AI systems in VR, research in this area has been limited. While various studies have investigated the use of VR for visualizing AI systems, a comprehensive review or meta-analysis has yet to consolidate the present state of knowledge in this field.

This comprehensive literature study is intended to give an in-depth overview of the present state of knowledge about the viability of virtual reality technology for visualizing AI systems. We anticipate that the findings of this review will offer valuable insights that can guide future research in the domain of VR-based AI visualization. The review highlights the potential and limitations of existing VR engines and visualization methods, thereby identifying key areas of focus for further development and innovation. This study provides a basis for understanding the challenges currently faced in this field, such as handling complex data and managing cognitive load within VR environments. It also underscores the need for further exploration into the ethical implications of integrating VR and AI technologies. Thus, it presents a roadmap to shape the evolution of AI visualization tools, encouraging the development of more intuitive, user-friendly, and ethically responsible solutions.

The main objectives of this review are to synthesize the findings from the literature, identify major themes and trends in the field, and offer suggestions for further investigation. In order to achieve these objectives, the following research questions are proposed:

\begin{itemize}
\item[\textbf{RQ1:}] What are the most effective visualization techniques and VR engines for visualizing AI systems?
\item[\textbf{RQ2:}] What are the benefits and limitations of using VR for visualizing AI systems compared to other visualization techniques?
\item[\textbf{RQ3:}] What gaps or limitations exist in the literature on the visualization of AI systems in VR, and what future research is needed to address these gaps?
\end{itemize}

By addressing these research questions, we aim to contribute to the existing knowledge on utilizing virtual reality technology in the visualization of AI systems and present an up-to-date overview of the relevant literature.

This paper is organized as follows: Section~\ref{sec:methods} begins with a comprehensive overview of the research methods employed. This encompasses the search strategy adopted, inclusion and exclusion criteria applied, study selection process, and methods of data extraction and synthesis, all focusing on the literature related to AI visualization in VR. Following this, Section~\ref{sec:resultsofsyntheses} elaborates on the findings of the review, examining the array of VR techniques and AI systems in use, as well as the applied visualization and interaction techniques. It also discusses the potential applications and implementation challenges that VR technology faces in AI data visualization. Finally, Section~\ref{sec:discussionandconclusions} presents a thoughtful discussion and conclusion on the existing state of the field, highlighting the dynamic interplay of tools, methods, and techniques in use, potential research gaps to be filled, challenges to be overcome, and the importance of ethical considerations. This section wraps up the paper with recommendations for future research to propel the field of VR in AI visualization further.

\section{Methods}
\label{sec:methods}

This comprehensive review examined research published in English in peer-reviewed journals or conference proceedings on the visualization of AI systems in VR but eliminated studies that still need to meet these requirements. By the type of AI system represented in VR, studies were categorized for synthesis in order to find common themes and trends and provide a thorough overview of the current state of the art.
\subsection{Search Strategy}\label{subsec:searchstrategy}

An exhaustive search was conducted in this comprehensive literature review to identify relevant studies on visualizing AI systems in virtual reality. The following databases were searched: IEEE Xplore, Scopus, Web of Science, Springer, and Google Scholar. The search strategy included relevant keywords and subject headings related to visualization techniques and virtual reality AI systems,\footnote{The specific search query was: \texttt{("virtual reality" OR "VR") AND ("artificial intelligence" OR "AI") AND ("visualization" OR "visualisation" OR "display" OR "mapping" OR "interpret*" OR "explain*")}}
and was limited to articles published in English, without a time restriction. Additionally, reference lists of identified articles were manually scanned for potential additional studies. The most recent results were obtained through a search performed on January~8, 2023. All identified articles were imported into Mendeley's reference management software for further analysis and screening.

\subsection{Inclusion and Exclusion Criteria}
Studies were included in this comprehensive review if they met the following criteria:

\begin{itemize}
\item Published in a peer-reviewed journal or conference proceeding.
\item Focused on the visualization of AI systems in virtual reality.
\item Provided empirical data, such as case studies, experiments, or user evaluations.
\end{itemize}

Studies were excluded if they:

\begin{itemize}
\item Were not published in English.
\item Focused solely on AI systems or virtual reality, but did not specifically address the visualization of AI systems within VR environments.
\item Lacked sufficient methodological detail.
\end{itemize}

\subsection{Study Selection} \label{subsec:studyselection1}
\begin{figure}[ht]
  \centering
  \tikzstyle{startstop} = [rectangle, rounded corners, minimum width=3cm, minimum height=1cm,text centered, draw=black, fill=red!30]
  \tikzstyle{process} = [rectangle, minimum width=3cm, minimum height=1cm, text centered, draw=black, fill=orange!30]
  \tikzstyle{decision} = [diamond, minimum width=3cm, minimum height=1cm, text centered, draw=black, fill=green!30]
  \tikzstyle{arrow} = [thick,->,>=stealth]
  \begin{tikzpicture}[node distance=1.7cm,scale = 0.7, every node/.style={scale=0.7}]
    \node (search) [process] {940 records identified through database searching};
    \node (screen) [process, below of=search] {940 records screened at title/abstract level};
    \node (exclude1) [process, right of=screen, xshift=6cm] {915 records excluded};
    \node (review) [process, below of=screen] {\Ntotal{} works screened at full text level};
    
    \node (excluded) [process, right of=review, xshift=5.2cm] {\Nexcl{} works excluded};
    \node (analysis) [process, below of=review] {\Nincl{} works included};
    \draw [arrow] (search) -- (screen);
    \draw [arrow] (screen) -- (exclude1);
    \draw [arrow] (screen) -- (review);
    \draw [arrow] (review) -- (excluded);
    \draw [arrow] (review) -- (analysis);
    
  \end{tikzpicture}
  \caption{Flow diagram of the literature search and selection process.}
  \label{fig:flow}
\end{figure}
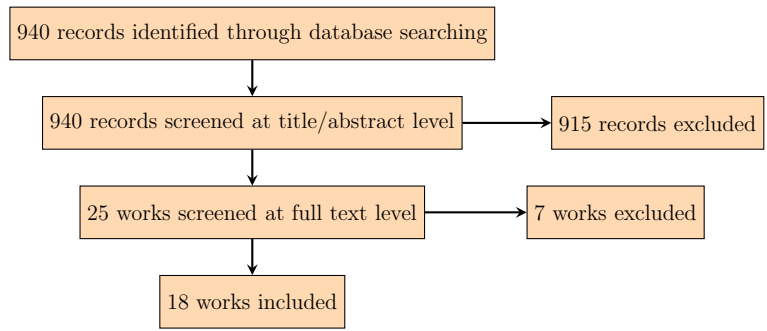

Figure~\ref{fig:flow} presents a flow diagram illustrating the literature search and selection process of this comprehensive review. Guided by a well-defined research strategy, we aimed to ensure the exhaustiveness and relevance of the included studies. The process began with the identification of 940 records from an extensive search of multiple databases. The Publish or Perish tool was employed as an additional filter at this stage to refine search results and ensure the quality of the selected studies \citep{Harzing2007}. This tool enabled a more precise assessment of the articles, leading to the inclusion of studies that held significant influence in the field and strong relevance to the research questions. Consequently, 915 records were disregarded due to irrelevance or non-compliance with inclusion requirements.

The remaining \Ntotal{} works underwent a full-text assessment to determine their suitability for the review. Subsequently, \Nexcl{} works were excluded for various reasons, including insufficient data, a lack of focus on visualization of AI systems in virtual reality, or other criteria-based factors. Ultimately, the final research included \Nincl{} papers, laying a robust foundation for synthesizing and assessing the state of knowledge on the visualization of AI systems in virtual reality.

\subsection{Data Extraction and Synthesis}\label{subsec:dataextractionandsynthesis}

We organized and summarized the included studies in detail in Table~\ref{table:studies} for analysis.

The synthesis process involved extracting key information from each study, such as reference, publication type, year of publication, VR engine/framework, features, and code availability. Each study is concisely outlined in the table, highlighting its primary goals and outcomes. The review's objective is to discern recurring patterns, shared aspects, and distinctions among the studies, thereby providing in-depth insight into the current state of knowledge regarding virtual reality's role in AI system visualization.

The table encompasses a diverse range of studies, including conference proceedings, journal articles, and tool development projects, all focusing on the visualization of AI systems in virtual reality. Most studies utilize widely-known VR engines, such as Unity and Unreal Engine, for frontend visualization, and backend machine learning frameworks like TensorFlow, PyTorch, and Caffe2. These studies explore features like network architecture, layer design, feature maps, and user interaction. Some also provide code availability, fostering further exploration and development by the research community.

By compiling information from these studies, the review presents a full overview of current research on the visualization of AI systems in virtual reality. It points out key advances and identifies areas requiring more research, ensuring a robust and in-depth understanding of the topic and laying the groundwork for future investigations.

\subsection{Study}\label{subsec:study}

The characteristics of each study, such as reference, publication type, year, features, VR engine/framework,  and code availability, have been thoroughly analyzed and are presented in Table~\ref{table:studies}. The table offers a comprehensive overview of the current research on the visualization of AI systems in virtual reality. By examining the characteristics of these studies, we aim to identify common themes, trends, and areas where further research is needed. This organized presentation of study characteristics in Table~\ref{table:studies} allows for a clear understanding and comparison of the key aspects of the included works, facilitating a comprehensive synthesis of the existing knowledge in this field.

During the study selection process, we identified several studies that appeared to meet the inclusion criteria but were eventually excluded:
\begin{enumerate}
\item \citet{Bock2018}---Excluded due to similarity to \citet{Schreiber2019}.
\item \citet{Wohlan2019}---Excluded due to significant overlap in content and authorship with \citet{Schreiber2019}.
\item \citet{prb2012}---Excluded due to its focus on visual data mining and representation of data and symbolic knowledge within VR spaces for classification tasks, rather than AI system visualization in VR.
\item \citet{Lamotte2010}---Excluded because it focused on the simulation of autonomous and intelligent vehicles in virtual reality, rather than on the visualization of AI systems in VR.
\item \citet{Meftah2020}---Excluded as its main focus was on creating a virtual simulation environment for autonomous vehicles to learn obstacle avoidance using deep learning, rather than directly addressing the visualization of AI systems in VR.
\item \citet{KALATIAN2021}---Excluded as it involves virtual reality and deep learning, its primary focus is on pedestrian crossing behavior in the presence of automated vehicles rather than on the visualization of AI systems in VR.
\item \citet{kundu2023vrlens}---Excluded as the paper is related to the broader field of AI and virtual reality; it does not specifically focus on the visualization of AI systems in VR. Instead, it presents a framework for detecting cybersickness in VR environments using machine learning models and explainable AI techniques. The goal of this research is to detect, analyze, and mitigate cybersickness in real-time for standalone VR headsets.
\end{enumerate}
These studies were excluded to ensure a complete and unbiased perspective on the topic in our comprehensive review.

\section{Results of Syntheses}\label{sec:resultsofsyntheses}

\subsection{Virtual Reality and AI Systems}

Different combinations of VR techniques and AI systems have been used in the study of AI visualization in virtual reality.

For example, \citet{Schreiber2019} utilized the Unreal Engine for creating 3D scenes dedicated to neural network visualization. They used TensorFlow as their AI system to provide detailed information for neural networks and permit the adjustment of training parameters. Similarly, \citet{Linse2022} employed the Unity game engine with OpenXR support for immersive visualization and interaction with convolutional neural networks (CNNs). Their AI system of choice was PyTorch, a deep learning framework designed to interconnect with Unity for dynamic visualization and interaction with custom CNN architectures.

Unity, as a gaming engine, was a common choice among researchers. For instance, \citet{Aamir2022} used Unity to create a virtual reality environment for visualizing deep neural networks built using the Caffe framework. \citet{Bibbo2022} and \citet{Nagasaka2021} both utilized Unity in their research, illustrating its dual functionality. They used Unity for creating virtual reality environments and developing deep convolutional neural network models. Furthermore, they employed Unity Barracuda to define these neural network models, demonstrating that Unity can also serve as a host for AI systems. \citet{Lyu2021} paired Unity with the Oculus Quest VR headset and used TensorFlow and Keras machine learning frameworks to define a CNN model for image classification.

\citet{Zhang2021} and \citet{Meissler2019} both used VR technology, with the latter also leveraging Unity and Steam VR, for deep learning model development. Their chosen AI system was deep learning neural networks defined using TensorFlow and Keras. \citet{Queck2022} and \citet{NARAHA2021} also relied on Unity for virtual reality techniques, with the latter using TensorFlow as the AI system.

In some studies, the focus was primarily on the VR techniques utilized, with less emphasis placed on the specific AI systems implemented. For instance, \citet{Bellgardt2020} leveraged Open VR, which supports a range of consumer head-mounted displays (HMDs) including Oculus Rift and HTC Vive, yet they did not detail the AI systems used. Similarly, \citet{Donalek2014} employed an array of VR tools including an Oculus Rift headset, a Leap Motion '3D mouse', and a Microsoft Kinect sensor, but did not disclose the AI systems integrated with these tools. In another instance, \citet{Bobek2021} mentioned the use of explainable AI and clustering algorithms, but they did not specify the exact software libraries or frameworks employed, emphasizing the variance in the level of detail provided across different studies. In summary, a range of combinations of virtual reality techniques and AI systems are evident in the literature. Unity stands out as a common choice for VR, with TensorFlow being frequently paired as the AI system. This diversity underscores the flexibility and adaptability of these technologies in visualizing AI systems in virtual reality.

Beyond the specific visualization of AI systems in VR, broader AI applications in VR have also been extensively reviewed. For instance, \citep{Oliveira2023} offers a comprehensive review of AI applications in VR, providing a wider perspective on how these two technologies can be combined to create interactive and immersive experiences.

\subsection{Visualization and Interaction Techniques}
Visualizing and interacting with complex AI models is crucial for understanding, interpreting, and explaining them. A variety of techniques are used to this effect, as evidenced by the following literature.

Interactive 3D visualization is a recurring technique used for exploring deep learning network layers at various levels of detail. \citet{Schreiber2019} utilized this approach, enabling users to interact with each neuron in the network. Similarly, \citet{Linse2022} used 3D rendering of Convolutional Neural Networks (CNNs), representing the computational graph as a connected conveyor system and optimizing the rendering of large architectures. They also allowed users to move, scale, and interact with CNN layers, offering a display of weight distributions, classification results, and feature visualizations. Expanding on these concepts, \citet{Hisham2022} delve into the merger of intricate networks and VR technology, creating interactive three-dimensional representations of large-scale datasets. Their approach is applied to a range of network types, including gene-gene interaction networks, social networks, and neural networks, demonstrating the broad applicability of these visualization techniques.

Visualization techniques can also involve more specialized approaches. For instance, \citet{Aamir2022} utilized interpretation modules, occlusion analysis, and virtual walkthroughs of network layers, with real-time manipulation of input images for interaction. \citet{Lyu2021} employed 3D force-directed graphs and real-time color changes to represent neural network parameters, with direct manipulation and sonification for auditory feedback as interaction techniques.

Certain studies use more straightforward visual representations, such as \citet{Bibbo2022}, who rendered 2D images with the gray colormap using matplotlib, though interaction techniques were not explicitly discussed. \citet{Zhang2021} used a direct physical approach, visualizing interactions by moving concrete objects with hands and showcasing real-time test dataset results.

\citet{Meissler2019} and \citet{Queck2022} presented network architecture, filters, and feature maps, with user interaction as the interaction technique. \citet{VanHorn2022} visualized feature maps, filter responses, and saliency maps in 2D layers, while offering menu navigation, grabbing and moving objects, and selecting layer properties as interaction techniques.

\citet{Bellgardt2020} utilized an immersive node-link visualization based on VR, with neurons spatially arranged in circles, and interactive elements controlled through the spatial input devices included with the Head-Mounted Display (HMD). They also included a virtual travel technique (flying) for user positioning.

In cases such as \citet{Bobek2021}, immersive parallel coordinates plots were used for visualization, although interaction techniques were not specified. \citet{Donalek2014} used immersive visualizations with SL Linden Scripting Language (LSL) and OpenSim, employing natural interaction techniques with commercial hardware.

Finally, \citet{Nagasaka2021} combined 3D visualization methods provided by TensorSpace.js and NeuralVis with conventional 2D visualization techniques such as Grad-CAM, while employing the Mixed Reality Toolkit and DXR for interaction.

In summary, a broad range of visualization and interaction techniques have been employed to illuminate and explore AI models. These methods offer different ways of understanding and interacting with AI systems, enriching our ability to interpret and utilize these complex models.

\subsection{Application and Implementation}

The utilization of VR technology for data visualization has vast potential, yet it is not devoid of challenges. For example, the complexity and high dimensionality of modern datasets can often make visualization difficult \citep{Korkut2023, Donalek2014}. Additionally, issues surrounding the scalability and adaptability of these visualization techniques to accommodate larger, more intricate architectures are prevalent \citep{Linse2022}.

Moreover, the cognitive load brought about by the complexity of the data and the VR environment itself is another concern. This has led to the call for more intuitive and user-friendly visualization tools to simplify the learning process and boost understanding \citep{Meissler2019, Lyu2021, Bibbo2022, Queck2022}.

Moreover, the limitations of current tools and technologies present considerable technical challenges. For example, \citet{Bibbo2022} pointed out the lack of support for large datasets in VR platforms and the limited resolution of existing VR display technology. The need for high-quality graphics and processing power, along with the potential for motion sickness in VR environments, create substantial obstacles \citep{Sharma2016}. \citep{Linse2022} say that in sensitive applications such as medicine and law enforcement, black-box systems such as neural networks will need to become more transparent in order to comply with regulations and earn the public's trust. This is compounded by the difficulties in understanding and representing the depth of images and models, as noted by \citep{Oliveira2023}. These intersecting technical and ethical considerations necessitate careful attention and innovative problem-solving in the continued development and implementation of VR visualization tools for AI systems.

A recent study by \citet{Hisham2022} provides a practical example of VR-based model visualization in the context of gene-gene interactions between human sex chromosomes and other human chromosomes. The authors detail the entire process of model development, from data collection to the final display of the model in a VR environment. The study also outlines specific software and hardware requirements, including the use of Python libraries, Unity 3D software, and Meta Quest 2 hardware. Despite the technical challenges involved, the study underscores the value of VR technology in enhancing the visualization and interpretation of complex network data.

Notwithstanding these challenges, promising applications of VR for data visualization have been demonstrated across various domains. For instance, VR has proven effective in interpreting and understanding complex AI structures, such as deep learning models and convolutional neural networks \citep{Schreiber2019, Linse2022, Aamir2022, Meissler2019, Queck2022}. The application of VR in these areas offers potential enhancements in explainability and interpretability of complex AI and machine learning systems \citep{Schreiber2019, Aamir2022}.

\section{Discussion and Conclusions}\label{sec:discussionandconclusions}

The synthesis of the literature reveals a dynamic interplay of tools, methods, and approaches in utilizing VR for the visualization of AI systems. The widespread adoption of versatile VR engines like Unreal Engine and Unity underlines the necessity for flexible and universally embraced foundations. These engines are compatible with an array of AI systems, including but not limited to ML platforms like TensorFlow, Caffe2, and PyTorch, showcasing their capacity to integrate with diverse technologies seamlessly. Furthermore, the capability of these systems to interface with a variety of VR hardware, including Oculus, HTC VR headsets, and numerous Head Mounted Displays, underscores the promising role of immersive technologies in creating an interactive and engaging environment for AI visualization.

The assortment of visualization and interaction techniques in the literature reflects the importance of intuitive and engaging methods for representing complex AI models. Interactive 3D visualization, immersive node-link visualization, and other specialized approaches illustrate the richness of techniques currently employed. However, the diversity of these methods underscores a potential research gap, as the lack of standardization could hinder collaborative efforts and delay the development of best practices. To address this, further comparative studies or meta-analyses might prove beneficial to identify the most effective combinations.

Despite the widespread application of VR in AI visualization across various domains, significant challenges persist. These include the high dimensionality and complexity of modern datasets, the cognitive load imposed by the VR environment, and the limitations of current tools and technologies. Thus, another research gap emerges: the need for more user-friendly, intuitive, and adaptable tools that can accommodate complex and high-dimensional AI models.

Moreover, the increasing integration of these technologies into our everyday lives necessitates a focus on ethical considerations. Several studies highlight the need for responsible and considerate use of AI and VR technologies, revealing another research gap that calls for more research into ethical guidelines or principles for this field.

Therefore, future research efforts must continue to innovate visualization and interaction techniques, address identified challenges, consider ethical implications, and fill the existing research gaps. Moreover, it is crucial that subsequent studies continue exploring the potential of VR for AI visualization across diverse application domains, while seeking to reduce the cognitive load and enhance user experience.

In conclusion, the application of VR in AI visualization is an evolving field marked by significant potential and substantial challenges. The research reviewed here provides a solid foundation for future studies, pushing forward a deeper understanding and more effective use of AI systems through immersive technologies. Nonetheless, substantial gaps exist in the current research that need to be addressed to further advance this promising field.

\section*{Funding}
This work is supported by Science Foundation Ireland, under Grant number 20/FFP-P/8853.

\nocite{*}

\bibliographystyle{plainnat}
\bibliography{references}

\appendix
\newcounter{rowcount}
\setcounter{rowcount}{0}
\afterpage{\clearpage
\begin{landscape}
  \section{Additional Data}
  \begin{scriptsize}
    \begin{longtable}{l >{\raggedright}p{7em} >{\raggedright}p{6em} >{\raggedright}p{2.cm} l  >{\centering\arraybackslash}p{7em} >{\centering\arraybackslash}p{19em} >{\centering\arraybackslash}p{5.5em}}
      \toprule
      \textbf{\#} & \textbf{Ref} & \textbf{Publication Venue} & \textbf{Type} & \textbf{Year}  & \textbf{VR Engine/ Framework} & \textbf{Features} & \textbf{Code \mbox{Availability}}\\\midrule\endhead
      \stepcounter{rowcount}0\arabic{rowcount}& \citet{Schreiber2019} & Conference & Tool Development & 2019  & Unreal
      Engine/ TensorFlow & Network architecture & --- \\*\cmidrule(l){3-8}
      \multicolumn{2}{ l}{\textbf{Summary}}&\multicolumn{6}{p{18.cm}}{The paper describes an application aimed at enhancing the comprehensibility of neural networks through interactive 3D visualization. The app allows users to visualize the layers of a convolutional neural network and observe the classification process, providing greater transparency and opacity of AI systems for both experts and non-experts. The prototype will be improved based on feedback and evaluation, with plans to support additional model types and data formats, as well as integration with augmented reality headsets. Future enhancements will include the addition of user interfaces for displaying neuron results, and support for standard exchange formats such as ONNX.}  \\
      \midrule
      \stepcounter{rowcount}0\arabic{rowcount}& \citet{Linse2022} & Journal & Tool Development, Research & 2022  & Unity/ PyTorch & Network architecture, Layer design, Feature maps, User interaction & Available \\*\cmidrule(l){3-8}
      \multicolumn{2}{ l}{\textbf{Summary}}&\multicolumn{6}{p{18.cm}}{This work introduces an open-source software for the immersive visualization of popular CNN architectures using Python and the Unity game engine, allowing users to freely navigate a 3D environment in desktop or VR mode. The software offers feature maps, activation histograms, weight histograms, and feature visualizations to facilitate a greater understanding of CNNs. The authors address the issue of making the visualization of large-scale models feasible in VR by developing a Pytorch module that enables the optimized visualization of nearly any computational graph, including branches and joints, in Unity. This software is made for both experienced developers and researchers, as well as those who are new to the field of deep learning.

        In a use case study, the authors trained the architectures CovidResNet and CovidDenseNet using three distinct training strategies on the Caltech101 and SARS-CoV-2 datasets to produce models with varying generalization abilities. Using visualization software, the authors determined that CNNs memorized images based on high-frequency patterns and proposed new measures to make it more difficult for the network to memorize images.

        The ability to visualize popular, cutting-edge architectures raises new issues for future research, such as considering machine learning problems with 3D input data and visualizing statistical variations in network characteristics. The authors propose clustering the channels based on the cross-correlation of their filter outputs in order to further improve the presentation of the feature space and to illustrate the semantic connections between channels.}  \\
      \midrule
      
      \begin{minipage}{0.0\linewidth}
        \begin{tabular}{l >{\raggedright}p{7em} >{\raggedright}p{6em} >{\raggedright}p{2.cm} l  >{\centering\arraybackslash}p{7em} >{\centering\arraybackslash}p{19em} >{\centering\arraybackslash}p{5.5em}}
        \stepcounter{rowcount}0\arabic{rowcount} & \citet{Sharma2016} & Journal & Survey & 2016  & --- & --- & ---\\*\cmidrule(l){3-8}
      \multicolumn{2}{ l}{\textbf{Summary}}&\multicolumn{6}{p{18.cm}}{The article explores the potential of integrating ANNs with VR to create immersive and interactive experiences. The primary objective of the review is to investigate the possible benefits of combining ANN and VR, discussing various algorithms and training functions related to ANN and their role in solving VR development and interfacing problems. The authors address the importance of data visualization in VR, highlighting how VR can enable effective data interaction and manipulation. They discuss the use of ANN for creating VR spaces and dimensionality reduction techniques, such as clustering and neural networks, which can enhance the efficiency of VR systems.
        The paper also presents several applications of ANN in VR, including facial expression detection, human body tracking, face detection, data visualization, and speech recognition. The authors argue that the integration of ANN and VR can lead to the development of powerful systems capable of a wide range of applications, creating intelligent virtual environments.

        In conclusion, the authors emphasize the potential benefits of combining ANN and VR technologies, and how their synthesis can contribute to more responsive and stimulated tracking and analysis of events in various fields, ultimately transforming existing practices and conventions.}
      \end{tabular}
    \end{minipage}\\
    \midrule
        
    \begin{minipage}{0.0\linewidth}
      \begin{tabular}{l >{\raggedright}p{7em} >{\raggedright}p{6em} >{\raggedright}p{2.cm} l  >{\centering\arraybackslash}p{7em} >{\centering\arraybackslash}p{19em} >{\centering\arraybackslash}p{5.5em}}
     \stepcounter{rowcount}0\arabic{rowcount} & \citet{Aamir2022} & Journal & Tool Development & 2022  & Unity/ Caffe2 & Network architecture, Layer design, Feature maps, User interaction& --- \\*\cmidrule(l){3-8}
      \multicolumn{2}{ l}{\textbf{Summary}}&\multicolumn{6}{p{18.cm}}{The paper discusses the use of VR technology for visualizing and interpreting DNNs. The authors developed a plugin for the Caffe framework in the Unity gaming engine to create and visualize a VR-based AlexNet architecture for an image classification task. The interactive model allows users to navigate through the network and select connections to understand the activity flow of particular neurons. An interpretation module based on the Shapley values algorithm was used to analyze the network's decisions. The authors suggest that VR-based visualization can provide a more immersive and accessible way to explore and interpret DNN models, which can help improve their decision-making processes. They also suggest possible future work, including developing a formal quantification method for interpreting network decisions.}\\
    \end{tabular}
  \end{minipage}\\
  \midrule
      
  \begin{minipage}{0.0\linewidth}
    \begin{tabular}{l >{\raggedright}p{7em} >{\raggedright}p{6em} >{\raggedright}p{2.cm} l  >{\centering\arraybackslash}p{7em} >{\centering\arraybackslash}p{19em} >{\centering\arraybackslash}p{5.5em}}
    \stepcounter{rowcount}0\arabic{rowcount} & \citet{Bibbo2022} & Journal & Research & 2022  & Unity/ Barracuda & --- & --- \\*\cmidrule(l){3-8}
      \multicolumn{2}{ l}{\textbf{Summary}}&\multicolumn{6}{p{18.cm}}{The article discusses using VR in deep learning and data visualization. The authors propose a VR platform using Unity for developing deep convolutional neural network models for image classification. The article describes the methodology used to create a CNN for recognizing human activities (HAR), which involves using the Barracuda package developed by Unity Labs and the ONNX format for transferring machine learning models. The article concludes that VR can be an effective tool for designing deep learning applications and is suitable for classifying images in various scientific sectors.}\\
    \end{tabular}
  \end{minipage}\\
  \midrule
      
  \begin{minipage}{0.0\linewidth}
    \begin{tabular}{l >{\raggedright}p{7em} >{\raggedright}p{6em} >{\raggedright}p{2.cm} l  >{\centering\arraybackslash}p{7em} >{\centering\arraybackslash}p{19em} >{\centering\arraybackslash}p{5.5em}}
   
     \stepcounter{rowcount}0\arabic{rowcount} & \citet{Lyu2021} & Conference & Tool Development & 2021  & Unity / Python & NN visualization: node-link approach, User interaction & --- \\*\cmidrule(l){3-8}
      \multicolumn{2}{ l}{\textbf{Summary}}&\multicolumn{6}{p{18.cm}}{This study introduces a virtual reality interface for interacting with artificial intelligence. THe interface lets users operate neural networks using virtual hands and offers audible feedback on the loss, accuracy, and hyperparameters in real time. The system's goal is to give a creative and user-friendly interface for interacting with AI, which may facilitate the understanding of the principles behind training neural networks. THe study suggests several future directions, including enhancing the system's pedagogical benefits, looking at designs for larger neural networks, and experimenting with new forms of sonification to enhance the user experience.}\\
    \end{tabular}
  \end{minipage}\\
  \midrule
      
  \begin{minipage}{0.0\linewidth}
    \begin{tabular}{l >{\raggedright}p{7em} >{\raggedright}p{6em} >{\raggedright}p{2.cm} l  >{\centering\arraybackslash}p{7em} >{\centering\arraybackslash}p{19em} >{\centering\arraybackslash}p{5.5em}}
   
      \stepcounter{rowcount}0\arabic{rowcount} & \citet{Zhang2021} & Journal & Research & 2021  & --- & --- & --- \\*\cmidrule(l){3-8}
      \multicolumn{2}{ l}{\textbf{Summary}}&\multicolumn{6}{p{18.cm}}{ This paper presents a deep learning model development environment based on VR technology for image classification. The proposed DNN environment allows users to build neural networks by moving concrete objects with their hands, automatically transforming these configurations into a trainable model and providing real-time test dataset results. The study highlights the significance of interactive technology in addressing challenges in understanding and analyzing neural networks. The system aims to bridge the gap between professionals in different disciplines, offering a new perspective on the model analysis and data interaction. The results demonstrate that the proposed DNN method outperforms traditional PCA and SVM methods in classifying environmental landscape images. The paper also discusses the implementation of real-time image processing algorithms on FPGAs, emphasizing the advantages of using large memory and embedded multipliers.}\\
    \end{tabular}
  \end{minipage}\\
  \midrule

      \stepcounter{rowcount}0\arabic{rowcount} & \citet{Meissler2019} & Conference & Tool Development & 2019  & Unity and STEAM VR/ TensorFlow & Network architecture, Feature maps, User interaction & Available \\*\cmidrule(l){3-8}
      \multicolumn{2}{ l}{\textbf{Summary}}&\multicolumn{6}{p{18.cm}}{This paper investigates the potential of VR for visualizing CNNs for individuals who are new to machine learning. Because neural networks are so complicated, the authors present a VR-based visualization method to help people who aren't experts understand how CNNs work. The study emphasizes the role of virtual reality in creating an immersive and engaging environment that increases learning motivation while reducing distractions.

        An exploratory study was conducted with 14 participants, most of whom had little knowledge of CNNs. The results showed that the VR visualization method was easy to use, helped people learn, and made them more interested in learning about CNNs. Participants found the VR environment more comfortable, fun, and exciting than traditional desktop visualizations. Some participants noted that the immersive nature of VR helped them focus better on the complex architecture of CNNs.

        The authors suggest doing a follow-up study comparing how well VR visualization works with traditional desktop visualisation. This would help us learn more about how immersive environments affect how well people learn. The study also gives ideas for improving VR visualisation by letting users move and resize individual visualization elements to make the structure fit their needs and preferences. This research adds to the growing interest in using VR in schools and workplaces to help people learn and understand difficult ideas.}\\\midrule
        \begin{minipage}{0.0\linewidth}
          \begin{tabular}{l >{\raggedright}p{7em} >{\raggedright}p{6em} >{\raggedright}p{2.cm} l  >{\centering\arraybackslash}p{7em} >{\centering\arraybackslash}p{19em} >{\centering\arraybackslash}p{5.5em}}
            
        \stepcounter{rowcount}0\arabic{rowcount} & \citet{Queck2022} & Conference & Tool Development & 2022  & Unity /--- & Network architecture, Feature maps, User interaction & --- \\*\cmidrule(l){3-8}
      \multicolumn{2}{ l}{\textbf{Summary}}&\multicolumn{6}{p{18.cm}}{The work aims to develop a VR application to visualize CNNs for machine learning beginners. The authors identified the need for an interactive and immersive visualization tool to aid in understanding CNNs, which can be complex and challenging for beginners. They conducted a proof-of-concept study with five participants and used the thinking-aloud method to evaluate the clarity of the VR environmen. The study found that the VR environment was intuitive and helpful for users to understand the CNN components. However, some users found it difficult to understand the relationship between the input layer and feature maps. In addition,  the visualization of the pooling layers did not stand out from the feature maps in their form of presentation. The authors plan to enhance the prototype based on the evaluation and conduct a user study to further evaluate its effectiveness.

        In conclusion, the study found that the VR environment was intuitive and helpful for users to understand the CNN components, but some areas need improvement. The authors plan to enhance the prototype and conduct further studies to evaluate its effectiveness. This work has important implications for machine learning, as it provides an interactive and immersive visualization tool to aid in understanding complex CNNs, which can lead to better learning outcomes for beginners.}
      \end{tabular}
    \end{minipage}\\
    \midrule
    \begin{minipage}{0.0\linewidth}
      \begin{tabular}{l >{\raggedright}p{7em} >{\raggedright}p{6em} >{\raggedright}p{2.cm} l  >{\centering\arraybackslash}p{7em} >{\centering\arraybackslash}p{19em} >{\centering\arraybackslash}p{5.5em}}

      \stepcounter{rowcount}\arabic{rowcount} & \citet{Bellgardt2020} & Conference & Tool Development & 2020  & Open VR/ --- & Network architecture, Feature maps, User interaction  & --- \\*\cmidrule(l){3-8}
      \multicolumn{2}{ l}{\textbf{Summary}}&\multicolumn{6}{p{18.cm}}{The article describes a new tool for visualizing ANNs  using node-link diagrams in immersive virtual reality. The tool is targeted towards machine learning experts and was evaluated through an expert review. The results of the review showed that the tool was perceived as helpful in a professional context, which supports the hypothesis that node-link visualization can improve the workflow of machine learning experts. While more evaluation is needed, the authors are optimistic that their visualization tool could be actively used by experts in the field.}\\      
    \end{tabular}
  \end{minipage}\\
  \midrule
  \begin{minipage}{0.0\linewidth}
    \begin{tabular}{l >{\raggedright}p{7em} >{\raggedright}p{6em} >{\raggedright}p{2.cm} l  >{\centering\arraybackslash}p{7em} >{\centering\arraybackslash}p{19em} >{\centering\arraybackslash}p{5.5em}}

      \stepcounter{rowcount}\arabic{rowcount} & \citet{NARAHA2021} & Conference & Research & 2020  & Unity/ TensorFlow & --- & --- \\*\cmidrule(l){3-8}
      \multicolumn{2}{ l}{\textbf{Summary}}&\multicolumn{6}{p{18.cm}}{This paper explores and proposes using visualization technology for deep learning and VR research projects, opening up new avenues for exploration in the field. The paper presents an outline of deep learning visualization research and case studies of deep learning visualization research using VR. The paper also identifies challenges that need to be addressed for effective visualization using VR, such as evaluation methods, high-quality operability, a wide range of customizability, scalability, and special support for beginners. Case studies and an overview of deep learning visualization research using virtual reality are presented in this paper.

        The research presented in this paper provides evidence that VR technology has the potential to enhance the way humans interact with deep learning models by providing insights into the processes of model construction and training. Significant difficulties are highlighted, and potential solutions are proposed, such as the use of new evaluation criteria and gamification to simplify deep learning for newcomers. Further, the paper argues that the proposed method's scalability and adaptability are crucial for experienced users and programmers. Overall, the paper shows the potential of virtual reality technology in deep learning research, and the authors suggest that implementing functions to customize filters and visualize results in real-time is essential for practical application.}\\
      \end{tabular}
    \end{minipage}\\
    \midrule
    \begin{minipage}{0.0\linewidth}
      \begin{tabular}{l >{\raggedright}p{7em} >{\raggedright}p{6em} >{\raggedright}p{2.cm} l  >{\centering\arraybackslash}p{7em} >{\centering\arraybackslash}p{19em} >{\centering\arraybackslash}p{5.5em}}

      \stepcounter{rowcount}\arabic{rowcount} & \citet{VanHorn2022} & Journal & Research & 2022  & Unity/ TensorFlow & Network architecture, Layer design, Feature maps, User interaction & Available \\*\cmidrule(l){3-8}
      \multicolumn{2}{ l}{\textbf{Summary}}&\multicolumn{6}{p{18.cm}}{In this study, the authors developed an interactive VR environment for constructing and analyzing deep learning models for biomedical image classification. The authors found that their proposed tool effectively enabled non-experts to understand and organize the structure of deep learning models and allowed users to build more accurate models and troubleshoot existing models faster when compared to a state-of-the-art drag-and-drop alternative. The authors also found that users enjoyed the experience in virtual reality significantly more, which they suggest can partially explain the higher objective scores, as positive emotions help with information retention. The authors argue that their interface achieved better outcomes through intuitive affordances for user actions, immersion, and ease of use.

        While introducing virtual reality presents challenges in educational and professional applications, such as cost, design, and physical limitations, the authors designed their VR environment to mitigate some risks. A more comprehensive VR platform could benefit expert use in the future, with improvements such as adding more layer types, developing new UI elements for fine-tuning layer properties, and enabling nonlinear model designs. The authors also suggest that improved visualization techniques and more explanatory elements could improve their tool's demonstrational benefit and interpretability. They conclude that future 3D/4D data visualization work can benefit from more straightforward navigation and a more accessible computational approach. Their proposed VR environment could be directly applied to cross-domain applications in biomedical image classification, providing sufficient benefits in understanding and prototype development.}\\
      \end{tabular}
    \end{minipage}\\
    \midrule
    \begin{minipage}{0.0\linewidth}
      \begin{tabular}{l >{\raggedright}p{7em} >{\raggedright}p{6em} >{\raggedright}p{2.cm} l  >{\centering\arraybackslash}p{7em} >{\centering\arraybackslash}p{19em} >{\centering\arraybackslash}p{5.5em}}

     \stepcounter{rowcount}\arabic{rowcount} & \citet{Bobek2021}  & Journal &---& 2021  &---& --- & --- \\*\cmidrule(l){3-8}
      \multicolumn{2}{ l}{\textbf{Summary}}&\multicolumn{6}{p{18.cm}}{The paper presents a refinement of the Immersive Parallel Coordinates Plots (IPCP) system for Virtual Reality (VR) and integrates data-science analytics, including explainable AI (XAI) methods, to enhance the visualization of multidimensional datasets in VR. The enhancements aim to automate part of the analytical work and assist users in pattern identification in complex datasets.

        The focus on visualization of multidimensional datasets and the integration of XAI methods in a VR environment aligns with the main focus of your systematic review, which is the visualization of AI systems in virtual reality.}\\
        
      \end{tabular}
    \end{minipage}\\
    \midrule
    \begin{minipage}{0.0\linewidth}
      \begin{tabular}{l >{\raggedright}p{7em} >{\raggedright}p{6em} >{\raggedright}p{2.cm} l  >{\centering\arraybackslash}p{7em} >{\centering\arraybackslash}p{19em} >{\centering\arraybackslash}p{5.5em}}
        
        \stepcounter{rowcount}\arabic{rowcount} & \citet{Korkut2023}  & Journal & Survey & 2023  & --- & --- & --- \\*\cmidrule(l){3-8}
      \multicolumn{2}{ l}{\textbf{Summary}}&\multicolumn{6}{p{18.cm}}{The paper presents a systematic literature review on visualization in VR. It analyzes various techniques used in different domains and their collaboration. The review found that there is a growing body of research on immersive visualizations across various problem domains. However, only a few studies focus on creating standard guidelines for VR, and each study either provides an individual framework or relies on traditional 2D visualizations. Game engines are widely used, but they are not suitable for critical scientific studies. The paper mentions two examples of AI visualization in VR: Gradient-weighted Class Activation Mapping (GradCAM) for Deep Reinforcement Learning (DRL) algorithms and Caffe2Unity for visualizing neural networks. The review highlights the need for further research and alternative approaches to address design challenges, develop standard guidelines for VR, and ensure the accuracy and effectiveness of 3D visualizations.}\\
    \end{tabular}
  \end{minipage}\\
  \midrule
  \begin{minipage}{0.0\linewidth}
    \begin{tabular}{l >{\raggedright}p{7em} >{\raggedright}p{6em} >{\raggedright}p{2.cm} l  >{\centering\arraybackslash}p{7em} >{\centering\arraybackslash}p{19em} >{\centering\arraybackslash}p{5.5em}}
      
      \stepcounter{rowcount}\arabic{rowcount} & \citet{Donalek2014}  & Conference & Tool Development & 2014  &---& --- & --- \\*\cmidrule(l){3-8}
      \multicolumn{2}{ l}{\textbf{Summary}}&\multicolumn{6}{p{18.cm}}{The authors leverage commercial software development for virtual environments such as video games or virtual worlds and focus on developing scientific data visualization tools within such environments. They develop immersive visualizations of highly dimensional data sets using general-purpose visualization techniques and scripts. They propose using natural interaction techniques in immersive virtual reality with inexpensive commercially developed hardware.}\\
    \end{tabular}
  \end{minipage}\\
  \midrule

      \stepcounter{rowcount}\arabic{rowcount} & \citet{Nagasaka2021}  & Conference & Tool Development & 2021  & Unity/ Unity Barracuda & --- & --- \\*\cmidrule(l){3-8}
      \multicolumn{2}{ l}{\textbf{Summary}}&\multicolumn{6}{p{18.cm}}{The paper presents an interactive visualization system for DL models in an immersive environment. The immersive environment allows for unlimited displays and visualization of high-dimensional data, making it possible to analyze data propagation through layers and compare multiple performance metrics. The proposed system addresses the challenge of limited display area in desktop environments and aims to make the analysis of complex DL models more accessible for non-experts. The prototype system received positive feedback from machine learning engineers, but they viewed the visualization technology as a unique introduction to the immersive environment. Future work includes improving system design, evaluating usability, comparing performance with existing desktop systems, and exploring the benefits of immersive visualization in DL model analysis. The ultimate goal is to develop hybrid systems that complement existing tools rather than replacing them.}\\\midrule
      \stepcounter{rowcount}\arabic{rowcount} & \citet{Hisham2022}  & Conference &Tool Development& 2022 & Unity/ Python & Network manipulation, Real-time evaluation, User interaction & --- \\*\cmidrule(l){3-8}
      \multicolumn{2}{ l}{\textbf{Summary}}&\multicolumn{6}{p{18.cm}}{This paper investigates the integration of complex networks and VR technology to create interactive 3D visualizations of large-scale data. This methodology is applied to various types of networks including gene-gene interaction networks, social networks, and neural networks. The authors illustrate the process of developing a VR-based visualization model using a biological dataset. They highlight both hardware and software requirements for implementing such VR visualizations. Examples from literature showcase the successful application of VR technology in understanding intricate relationships in these networks. The authors predict an increase in the use of VR technology for data visualization, particularly with the emergence of Metaverse concepts. As future work, they plan to develop a large-scale gene-gene interactions dataset and a VR interactive application to provide an efficient model for specialists to interact with large-scale data.}  \\\midrule
      \stepcounter{rowcount}\arabic{rowcount} & \citet{Oliveira2023}  & Journal & Survey & 2022 & --- & --- & --- \\*\cmidrule(l){3-8}
      \multicolumn{2}{ l}{\textbf{Summary}}&\multicolumn{6}{p{18.cm}}{This comprehensive literature review investigates the application of AI methods in VR solutions, given the scarcity of such studies. The analysis involved locating and evaluating relevant documents from various databases, with a particular focus on AI's contributions to VR applications. The study observed that machine learning, specifically in the subfields of neural networks, deep learning, and fuzzy logic, is the most prevalent AI technique employed in VR. The application of AI in VR revealed multiple advantages, including high algorithmic efficiency and precision, especially in human-machine interaction and intelligent robotics. The study also revealed numerous real-world application fields such as emotion interaction, education, agriculture, transport, and health. However, the review highlighted several limitations, such as high computational cost and dataset dependence. This paper emphasizes the potential for future research focusing on finding new VR applications incorporating AI technologies, alongside a stronger emphasis on AR.}  \\
       \bottomrule
      \caption{Included Studies}
      \label{table:studies}
    \end{longtable}
  \end{scriptsize}
\end{landscape}
\clearpage
}

\end{document}